\RequirePackage{lineno} 

\documentclass[12pt]{article}
\usepackage{setspace}
\usepackage[in]{fullpage}
\usepackage[numbers,sort&compress]{natbib} %
\usepackage{lineno}

\usepackage{graphicx}
\usepackage{amssymb}
\usepackage{bm}  

\begin{document}

\normalsize 

\title{\textbf{Pricing Using a Homogeneously Saturated Equation}}


\author{Daniel T. Cassidy \\ \\ Department of Engineering Physics, \\ McMaster University, \\ Hamilton, ON, Canada L8S 4L7 \\  \\ cassidy@mcmaster.ca \\}


\date{11 September 2011; revised 21 January 2013}

\maketitle

\begin{abstract} 
\normalsize

A homogeneously saturated equation for the time development of the price of a financial asset is presented and investigated for the pricing of European call options using noise that is distributed as a  Student's $\bm{t}-$distribution.  In the limit that the saturation parameter of the equation equals zero, the standard model of geometric motion for the price of an asset is obtained.  The homogeneously saturated equation for the price of an asset is similar to a simple equation for the output of a homogeneously broadened laser.  The homogeneously saturated equation tends to limit the range of returns and thus seems to be realistic.

\bigskip
Fits to {linear} returns obtained from the adjusted closing values for the S\&P 500 index were used to obtain best-fit parameters for Student's $\bm{t}-$distributions  {and for normal distributions}, and these fits were used to price options, and to compare approaches to modelling prices.  

\bigskip
This work has value in understanding the pricing of assets and of European call options.

\bigskip
\bigskip

keywords: homogeneously saturated; pricing; European call options; Student's $\bm{t}-$distribution;  {linear returns}
\end{abstract}



\linenumbers
\doublespacing

\section{Introduction} 
\label{intro}
A homogeneously saturated equation for the time development of the price of a financial asset is presented and investigated for the pricing of European call options using noise that follows Student's $\bm{t}-$distributions.  Parameters for the Student's $\bm{t}-$distributions were obtained by fitting to the {linear} returns calculated from the adjusted closing values of the S\&P 500 Index over the period of 3 January 1950 to 27 July 2011.  The homogeneously saturated equation for the price of an asset is similar to a simple equation for the output of a homogeneously broadened laser \cite{Cas2}.  In the limit that the saturation parameter $\beta$ of the homogeneously saturated equation equals zero, the standard solution of geometrical motion for the price of an asset is recovered.

The homogeneously saturated equation was obtained by constructing simple, coupled rate equations for the time development of the price of an asset and the supply of money.  These equations were solved in a steady state approximation to obtain an equation for the time development of the price of an asset.  The phenomenological approach is outlined in the Appendix and is presented in \cite{Cas4}.   {Analytic solutions to a homogeneously saturated model, which is coupled, non-linear differential equations, were originally presented in 1984 \cite{Cas2}.}

In the text that follows, $S(t)$ is the value of an asset at time $t$, $S_{{\rm o}} = S(0)$ is the value of the asset at $t=0$, $\alpha$ is a drift rate, $\sigma$ is a scale parameter, $f(t)$ is a {noise driving term and is a} stochastic process, and $\beta$ is a saturation parameter.  

In a Langevin approach, the equation for the time development of the value of an asset from a homogeneously saturated model is

\begin{equation}
{\frac{\rm d}{{\rm d} \, t}} S(t)~=~{\frac{\alpha \, S(t) ~+~\sigma \, S(t)\, f\, (t)}{1~+~\beta \, S(t)}} \,\,.
\end{equation}
In the development of the Eq. (1),  a reservoir for money was assumed and the noise was ascribed to fluctuations in the amount of money available to invest in the asset.  This is consistent with market microstructure studies \cite{Far, Smi, Tot, Dan}.  In these studies, it was revealed that the order book for the market is sparse, that price jumps occur when the orders are filled across gaps in the order book, that the stability of markets relies on a delicate balance between supply and demand, and that the need to store supply and demand to enable trading leads to structure in the pricing.  These studies show elements of a reservoir, of a  coupling between price and supply, and of fluctuations in the money available in the order book causing fluctuations in the price,  similar to the phenomenological elements that lead to a homogeneously broadened equation for the price of an asset.

The value of the asset at time $t$, $S(t)$, is found through solution of the {Langevin equation for the} homogeneously saturated {model, Eq. (1),} to be   
\begin{equation}
S(t) ~=~{\frac{S_{o} \, e^{\int_{0}^{t} \alpha \,\, +\,\, \sigma  f (\eta )\,\,{\rm d} \, \eta} }{e^{\,\beta \, (S(t)~- ~S_{o} )} }} ~=~{\frac{S_{o} \,e^{\,\alpha \, t\,\, +\,\, W(t)\,\,} }{e^{\,\beta \, (S(t)~-~S_{o} )} }} ~{\rm ,}
\end{equation}
a transcendental equation that can be solved for {\it S\/}({\it t\/}) given $\alpha$, $\beta$, {\it S$_{{\rm o }}$\/}, and $W(t)=\int_0^t \sigma f(\eta)\rm\,{d}\eta$.  The
simple dependence on the  {stochastic} process {\it W\/}({\it t\/}) makes simulation straight forward.   

By comparison, in a Langevin approach the standard model for the time development of the value of an asset is
\begin{equation}
{\frac{\rm d}{{\rm d} \, t}} S(t)~=~\alpha \, S(t)~+~\sigma \, S(t)\, f\, (t)
\end{equation}
with solution
\begin{equation}
S(t)~=~S_{o} \, \exp \int_{0}^{t} \left ( \alpha ~+~\sigma \, f\, (\eta ) \right ){\rm \,} \, {d}\eta ~=~S_{o}\, \exp \left ( \alpha t~+~W(t) \right )~.
\end{equation}

If the noise driving term $f(t)$ is a normally distributed process, then the solution to the standard model is geometrical Brownian motion. 

The standard model is obtained from the homogeneously saturated equation in the limit that the saturation parameter $\beta$ equals zero.

Numerous revert-to-mean models for the time development of the price of an asset,  of volatility, and of interest rates have been put forth and analyzed.   Anteneodo and Riera~\cite{Ant} and Wu et al.~\cite{Wu} have listed some of these models and have shown that these models can be incorporated in a single stochastic differential equation by choice of parameters in the single differential equation.  

It is doubtful that a simple revert-to-mean feature is of much benefit when the noise driving term is drawn from a  fat tailed distribution.  The standard model implicitly has a revert-to-mean feature;  a revert-to-mean value of zero, which is the mean value of the noise distribution and which is implicit in each model.  If the noise drives the output to a large value with a revert-to-mean of zero, then the noise will also drive the output to a large value when the revert-to-mean value is finite and non-zero.  Anteneodo and Riera~\cite{Ant} have shown that nonlinear additive-multiplicative processes are necessary to provide realistic descriptions of observations.

Smith et al.~\cite{Smi} used a rate equation for the density of the order book and non-linear feedback to investigate how prices depend on the rate of flow of orders.  The authors were able to explain the concavity of the price impact function, the existence of universal supply and demand functions, and the average daily spread.  Their model did not include coupled rate equations. 

The homogeneously saturated equation presented in this paper is based on phenomenological, coupled rate equations for the time development of a reservoir of money and of the price of an asset.  The rate equation for the reservoir of money has a revert-to-pumping-rate feature, where the pumping rate is the rate at which money flows into the reservoir to support the price.  The rate equation for the reservoir of money also has a stochastic driving term to account for fluctuations.  The steady state solution to the coupled equations has some interesting and realistic features.  The equation tends to produce a linear relationship between the input and output, and to limit the range of the output for a given input.  Both of these tendencies mean that the price of an asset is less likely than the standard model to wander off to infinity and that integrals required to price options can be evaluated for many fat-tailed distributions. 

The price of an European call option at time $T$ can be found from the arbitrage theorem~\cite{a1,a2,a3} as E$\{\max(S_T - K_T,0)\}$ where $S_T=S(T)$ is the {fair} price of the asset at time $T$,  $K_T$ is the strike price at time $T$, and E$\{\max(x,y)\}$ is the expectation of the maximum value of the set $\{x, y\}$ \cite{Cas1}.  {The constraint of a fair price for $S_T$ requires that $S(t)$ be a martingale.}  It is not necessary to solve a differential equation to find the price of an European call option \cite{r1, McM, Cas1}.  The arbitrage theorem approach gives the same answer as the Black-Scholes equation in the limit that the returns are normally distributed \cite{Cas1}.

In this paper the time development of the price of an asset and the pricing of European options based on a homogeneously saturated equation are considered.  Section 2 provides information on solutions to the homogeneously saturated equation and on simulations.  Section 3 provides fits to returns for the S\&P 500 data.  Section 4 provides information on pricing using the fits from Sec.~3. Section 5 is a conclusion and Sec.~6 is an Appendix that outlines a derivation of the homogeneously saturated equation.  The starting point for this paper is the solution to the homogeneously saturated equation.  The Appendix is not a rigorous derivation of the underlying equations.  The Appendix is meant as a phenomenological justification for the starting point for this paper, which is the solution to a homogeneously saturated equation for the time development of the price of an asset.  

{In this paper, time is represented by $t$ and the `\textit{t}' in the Student's $\bm{t-$}distribution is set in bold.}

\bigskip

\section{Homogeneously Saturated Equation}
Figure 1 is comprised of plots of the solution $S(x)$ to the homogeneously saturated equation{, Eq. (2),} as a function of the input  $x =\alpha t + W(t)$ for selected values of the saturation parameter $\beta$.  The shapes of the solutions for various values of $\beta$ are important in understanding the properties of the solution.  For small values of the saturation parameter $\beta$ the solution $S(x)$ appears to be exponential with $x$ whereas for large values of the saturation parameter the solution is essentially linear in $x$, with decreasing slope as $\beta$ increases.  The exponential increase with $x$ for small $\beta$ is consistent with the solution to the standard model, Eq. (4).

\begin{figure}[htbp] 
\centering
\includegraphics[scale = 0.7]{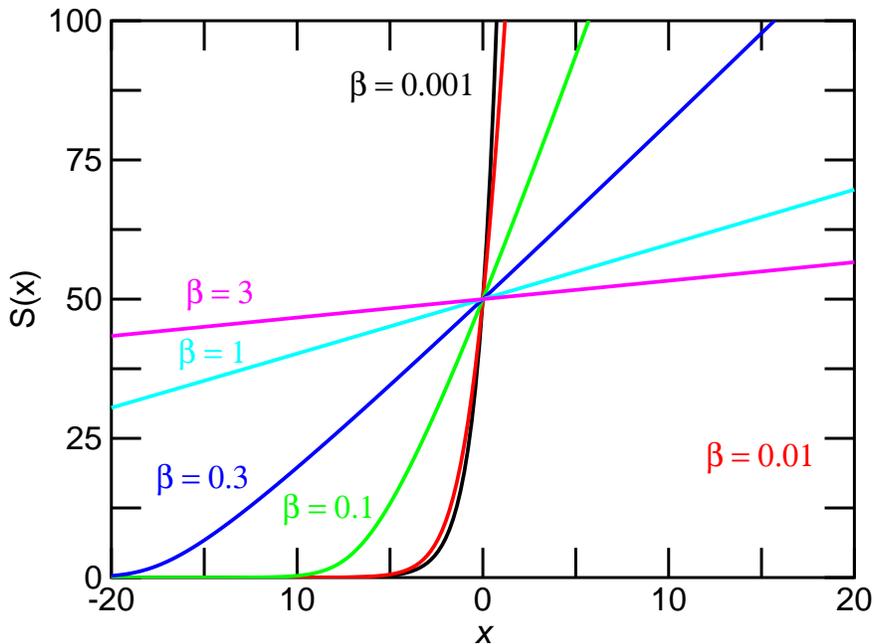}
\caption{{\normalsize Solutions $S(x)$ to the homogeneously saturated equation as a function of the input $x =\alpha t + W(t)$}}
\label{fig:fig1} 
\end{figure} 

\bigskip

Table~\ref{tbl1} gives descriptive statistics for solutions $S(x)$ {to Eq. (2)} for various values of the saturation parameter $\beta$ to the homogeneously saturated equation for 131072 random draws from a Student's $\bm{t}-$distribution with shape parameter $\nu = 3$ and for 131072 random draws from a normal distribution.  The draws from the normal and the $\nu=3$ Student's $\bm{t}-$distributions are the values for the sum of the drift and noise driving term, $x = \alpha t + W(t)$, in the equation for the price $S(x)${, Eq. (2)}.  The same set of random numbers were used for each value of $\beta$ but different sets were used for the normal and $\bm{t}-$distributions.  The scale parameters were adjusted, by multiplying the input noise by an appropriate factor,  for each $\beta$ to give the same standard deviation for $S(t)$.  This was done to enable comparison.  The standard deviation of the daily returns is observable.  The noise required to achieve the observed standard deviation is not known {\it a priori} in the homogeneously saturated equation. 

For the simulations reported in Table~\ref{tbl1}, {$S_T = S_o\, \exp(rT)$ was set equal to $50.00\,\exp(+0.03)$ $= 51.523$} and the standard deviation {of the solution to Eq. 2, $S(t)$,} was set to $S_T \times 0.3\times \sqrt{\nu/(\nu-2)}$, or $26.78$ for $S(t)$ found with the Student's $\bm{t}-$distribution driving noise and $15.46$ for $S(t)$ found from the normally distributed noise.  For a normal distribution, $\nu=\infty$, and $ \sqrt{\nu/(\nu-2)} = 1$.  The numbers were chosen to demonstrate pricing with $T=1$ year, the risk free rate $r=0.03$, the annualized volatility $=0.3$, and a very fat-tailed noise distribution with $\nu=3$ degrees of freedom (or a shape parameter $\nu=3$).  The Black Scholes formula sets{, under these conditions,} a price of $C_o = \$7.12$ for an European call option {with $K_T=\$ 49.00$}.  Also given in the Table~\ref{tbl1} are values from the Monte Carlo simulations for the prices of an European call option for the various values of $\beta$.  {The Black-Scholes price is given as point of reference only.  There are differences in approaches that lead to the  Black-Scholes equation and to the homogeneously saturated equation, and these differences make direct comparisons difficult.  See later in this Section for a discussion of noise rectification, of returns, and of volatility and standard deviation.}

\begin{table}
\caption{Descriptive statistics for $S(t)$ from a homogeneously saturated equation with Student's $\bm{t}$ and normally distributed noises.  {The costs $C_o$ of European call options with $S_o=50.0$, $K_T=49.0$, $T=1$ year, $r=0.03$, and volatility $=0.3$ are also given for the two noise sources.}}
\begin{tabular}{c c c c c c c} \hline
 & $\beta = 0.001$ & $\beta = 0.01$ & $\beta = 0.1$ & $\beta = 0.3$ & $\beta = 1.0$&$\beta = 3.0$ \\ \hline
$\bm{t}$ with $\nu=3 $ &  &  &  &  &  &\\ 
min & $0.02$ & $0.0$ & $0.0$ & $0.0$ & $0.0$ & $0.0$\\ 
average & 53.6 & 54.7 & 53.3 & 52.8 & 52.6 & 52.5 \\ 
max & 4809 & 1902 & 1138 & 1055 & 1023 & 1013 \\ 
kurtosis & 9211 & 334 & 55 & 42 & 38 & 37 \\ 
skewness & 67 & 9.3 & 3.1 & 2.6 & 2.4 & 2.3 \\ 
input scale & 0.172 & 0.410 & 2.07 & 5.64 & 18.1 & 53.6 \\ 
$C_o$ & \$5.91 & \$8.62 & \$10.04 & \$10.21 & \$10.27 & \$10.29 \\ \hline
normal &  &  &  &  &  &\\ 
min & $13$ & $9.6$ & $1.28$ & $0.0$ & $0.0$ & $0.0$\\ 
average &53.5 & 53.0 & 51.9 & 51.6 & 51.5 & 51.5 \\ 
max & 170 & 146 & 125 & 121 & 120 & 120 \\ 
kurtosis & 4.2 & 3.4 & 2.9 & 2.9 & 3.0& 3.0 \\ 
skewness & 0.83 & 0.56 & 0.16 & 0.07 & 0.02 & 0.005 \\ 
input scale & 0.300 & 0.451 & 1.87 & 4.97 & 15.8 & 46.7 \\ 
$C_o$ & \$7.03 & \$7.16 & \$7.28 & \$7.29 & \$7.29 & \$7.29 \\ \hline

\end{tabular}
\label{tbl1}
\end{table}

The 131072 samples for the input noise from the $\nu=3$ Student's $\bm{t}-$distribution had a mean of $7.72 \times 10^{-4}$, a standard deviation of $1.701 \approx \sqrt{\nu/(\nu-2)}=\sqrt{3}$, a minimum value of $-46.4$, a maximum value of 53.9, a skewness of 0.085, and a kurtosis of 39.9.  The 131072 samples for the input noise from the normal distribution had a mean of $-3.0 \times 10^{-3}$, a standard deviation of 1.000, a minimum value of $-4.67$, a maximum value of 4.39, a skewness of $-0.008$, and a kurtosis of 3.01.  For a Student's $\bm{t}-$distribution, which has support over $[-\infty, +\infty]$, the coefficient of skewness equals zero and exists for $\nu >3$, and the coefficient of kurtosis equals $3(\nu-2)/(\nu-4)$ and exists for $\nu>4$.  The skewness and kurtosis exist for all values of the shape parameter $\nu$ if the fat tails of the Student's $\bm{t}-$distribution are removed by truncation \cite{Cas5}.

Note that there is minimal noise rectification (i.e., the mean value for $S(t)$ does not have an $\exp(+\sigma^2/2)$ enhancement) for a heavily saturated system, and that the homogeneously saturated model shows a reduction of the maximum value for $S(t)$ as $\beta$ increases.  As $\beta$ increases, $C_o$, the price of a European option, approaches the Black Scholes value of \$7.12 for simulations that draw from a normal pdf.  {In the calculation of $C_o$, the mean and the standard deviation were forced through an iterative procedure to equal $S_o \exp(0.03)$ and $S_T \times 0.3\times \sqrt{\nu/(\nu-2)}$.  The stopping criterion for the iterative procedure was that the absolute value of the difference between the standard deviation for the solution to $S(t)$ and the target standard deviation was $< 0.0005$.}  Pricing with realistic inputs, rather than with input parameters chosen to demonstrate the differences, is given in Sec.~4.

The rows of Table~\ref{tbl1} that are labelled ``input scale" give the scale factors that were required to achieve the target standard deviation of $S(t)$.  The reason for the different input scales to obtain the target standard deviation for the output $S(x)$ can be observed from Fig. \ref{fig:fig1}.  For small $\beta$ and for $x = \alpha t + W(t) > 0$, $S(x)$ is a very steep function of $x$, and only small amplitudes for the input noise $x$ are required to give a large standard deviation for the output $S(t)$.  For large $\beta$ and $x$ such that $S(x) \gg 0$, $S(x)$ is, to a good approximation, a linear function of $x$ with a small slope, and large amplitudes for the input noise $x$ are required to achieve a large standard deviation for the output $S(x)$.   For $x$ such that $S(x) \gg 0$, the slope of $S(x)$ approaches $1/\beta$.  This means that the output is compressed for a heavily saturated system (i.e., for a system characterized by a large $\beta$).   Large changes in $x$ produce small changes in $S(x)$ for large $\beta$ and for $S(x) \gg 0$.

Figure~\ref{fig:fig1} can also  be used to explain the noise rectification properties.  For the normal data, the input noise (before the input scale is applied) is confined in the range $-4.67 \le x  \le 4.39$.  The  noise seldom pushes the solution $S(x)$ into the highly nonlinear region where $S(x) \approx 0$.  As a result, the pdf for $S(x)$ is approximately symmetric and centred around $x=0$; see the summary statistics for the normally distributed input of Table~\ref{tbl1}.  For the fat-tailed distribution, the input noise (before the input scaling is applied) is in the range $-46.4 \le x \le 53.9$, and the noise often pushes the solution $S(x)$ into the nonlinear region where $S(x) \approx 0$ and this distorts the pdf for $S(x)$, which leads to an increase of the mean value of the output.  For increasing $\beta$, $S(x)$ is an approximately linear function of $x$ over a greater range of $x$, and the contribution to the mean from the noise rectification drops.  For small values of the input noise (the shape and scale parameters of Table~\ref{tbl1} were chosen to highlight the effects), the noise rectification will be small or negligible, particularly as the transfer function for $S(x)$ becomes linear with increasing saturation parameter $\beta$.

\begin{figure}[htbp] 
\centering
\includegraphics[scale = 0.7]{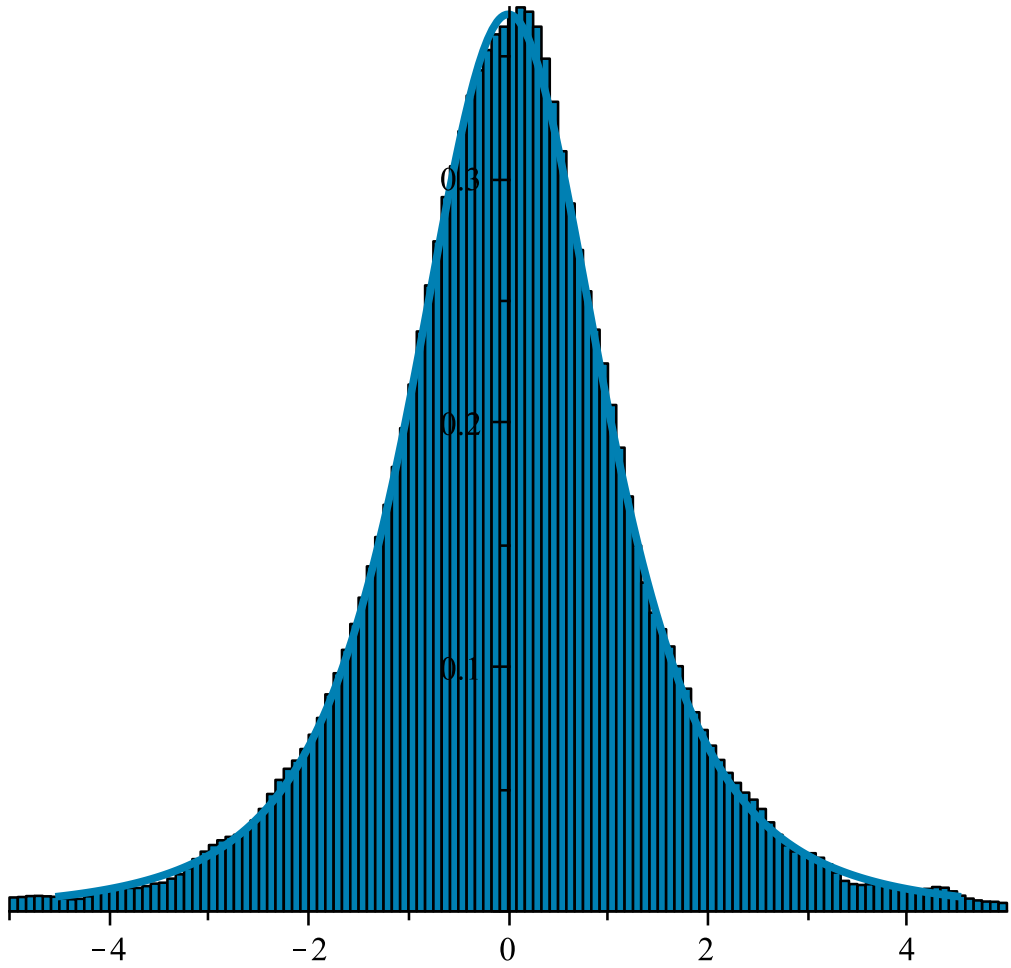} 
\caption{{\normalsize $\beta = 0.3$ histogram and Student's $\bm{t}-$distribution}}
\label{fig:fig6} 
\end{figure}

\begin{figure}[htbp] 
\centering
\includegraphics[scale = 0.7]{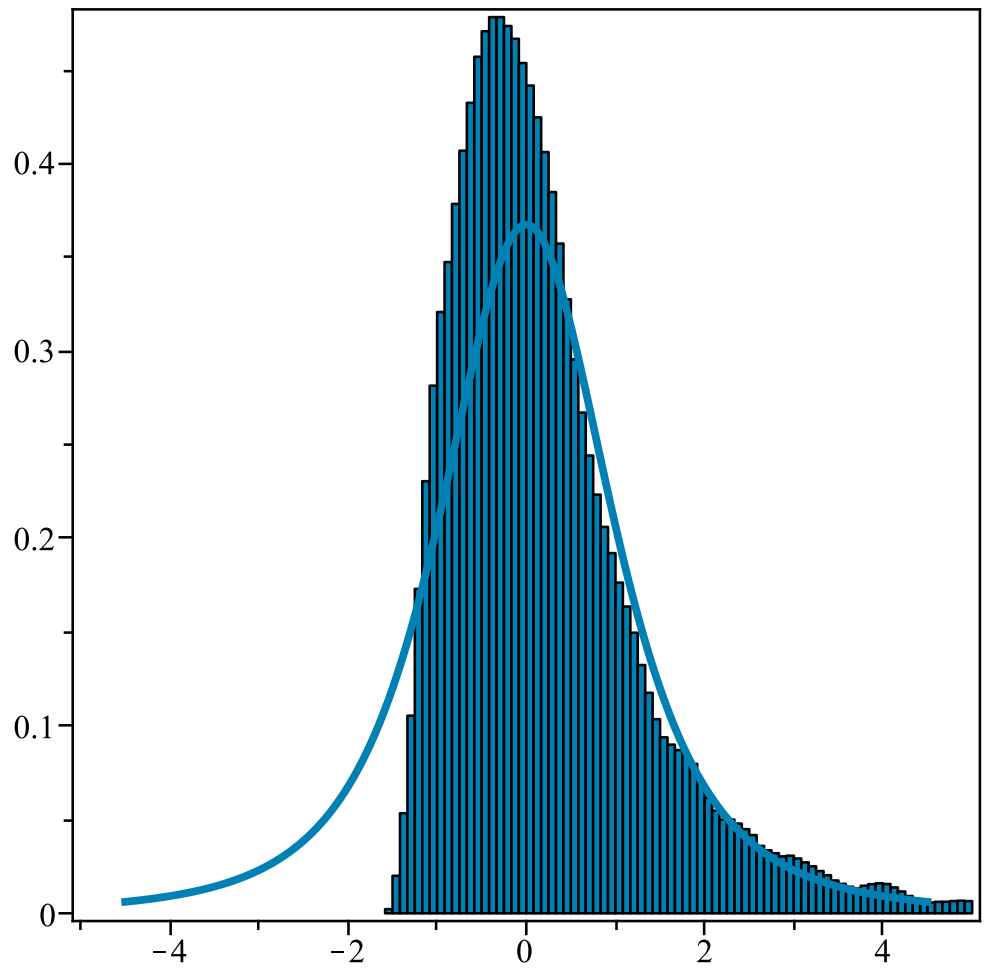} 
\caption{{\normalsize $\beta = 0.01$ histogram and Student's $\bm{t}-$distribution}}
\label{fig:fig5} 
\end{figure}

Figures~\ref{fig:fig6} and ~\ref{fig:fig5} are histograms of $S(x)$ for $\beta = 0.3$ and $0.01$ for 8192 draws from the same  $\nu=3$ Student's $\bm{t}-$distribution with a scale parameter equal to 0.3.  The solid curves are $\nu=3$ Student's $\bm{t}-$distributions with the standard deviations matched to the standard deviations of the solutions $S(x)$ for $\beta = 0.3$ and $0.01$.  The histograms show the asymmetry in the solution for small saturation parameter $\beta$ and the symmetry in the solution for large $\beta$ and small noise.  The mean of $S(x)-S_o$ for $\beta=0.01$ was 14.4 whereas the mean of $S(x)-S_o$ for $\beta=0.3$ was $-0.0024$.  For $\beta=0$, which is a log Student's $\bm{t}-$distribution, the mean value was $3.7 \times 10^5$ with a minimum value of 0.0 and a maximum value of $1.7 \times 10^9$ for the same noise driving terms for the 8192 simulations.  The homogeneously saturated equation limits the range of values and the noise rectification compared to the standard model (for which $\beta=0$).

\bigskip

\begin{figure}[htbp] 
\centering
\includegraphics[scale = 0.7]{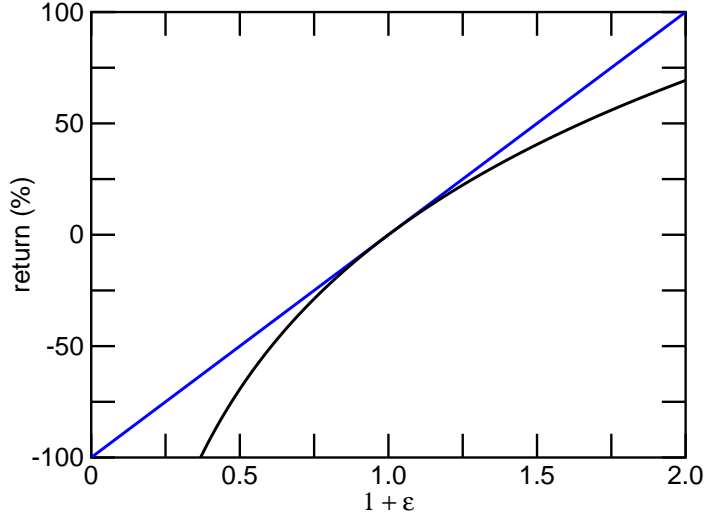} 
\caption{{\normalsize {Linear return (straight line) and logarithmic return (curve)} as a function of $S(t+1)/S(t) = 1 + \varepsilon$.   {The minimum possible return is when $S(t+1) = 0$, which  occurs for $1+\varepsilon=0$.}}}
\label{fig:fig3} 
\end{figure}

\bigskip
In this work the $n$-day per mille return $R_n(t)$ has been defined as $1000 \times (S(t+n)/S(t) - 1)$ where time $t$ and $n$ are measured in days.  The multiplicative factor of 1000 transforms the return into a per mille return.  The definition of the $n$-day return is in contrast to the more common $n$-day return of $\ln(S(t+n)/S(t))$.  Both definitions of the $n$-day return give the same answer for small deviations of $S(t+n)$ about $S(t)$, as can be observed in Fig.~\ref{fig:fig3}.  For small changes, $\ln(S(t+n)/S(t)) = \ln(S(t)(1+\varepsilon)/S(t))= \ln(1+\varepsilon) \approx \varepsilon -  \varepsilon^2/2 + \varepsilon^3/3$ and to first order, the logarithmic return is linear in $\varepsilon$.  There are several reasons for preferring the $n$-day linear return as defined here.  Returns are known to be fat tailed; the changes about $S(t)$ might not always be small.  The logarithmic form of the return permits returns that are $<-100 \%$, as shown in Fig.~\ref{fig:fig3}.  A return of $<-100 \%$ makes little sense.  The logarithmic return makes some sense in the standard model; $\ln(S(t)) = \alpha t + W(t)$ gives the sum of the noise $\sigma\,f(t)$ driving the process.  If $S(t)$ is distributed as log-normal, then the logarithm of $S(t)$ is normally distributed.   For a solution to a homogeneously saturated equation that is fully saturated, the transfer function is essentially linear -- the output is to a very good approximation a linear function of the input, as shown in Fig.~\ref{fig:fig1}.  A logarithmic return would then give in a homogeneously saturated approach a return that is the logarithm of the input noise, which would be acceptable (i.e., yield the same result as for the linear return $R_n(t)$) for small returns.  However, the definition $R_n(t)$ works for all magnitudes of changes and does not give returns $< -100 \%$, and thus seems preferable to the logarithmic definition.

\bigskip

In this work the volatility was taken as the scale parameter for the underlying distribution.  For a normal distribution, the scale parameter is the standard deviation.  For a Student's $\bm{t}-$distribution, the standard deviation is a function of the shape parameter $\nu$ times the scale parameter, and is not equal to the scale parameter.  A Student's $\bm{t}-$distribution with shape parameter $\nu$, scale parameter $b$, and location parameter (i.e., mean) equal to zero, can be written as 

\begin{equation} 
f_{\bm{t}}\;(\xi,\nu ,b )\;\mathrm{d}\xi\ =\ \frac{\Gamma ((\nu +1)/2)}{\sqrt{\pi \nu }\Gamma (\nu /2)}\left(\frac{1}{1\ +\ {\xi^{2}}/{(\nu b ^{2})}}\right)^{\frac{\nu
+1}{2}}\frac{\;\mathrm{d}\xi}{b }
\end{equation}
{where $f_{\bm{t}}(\xi,\nu ,b)\,\mathrm{d}\xi$ gives the probability of obtaining a value in the range $\xi$ to $\xi+\mathrm{d}\xi$ for a draw from a Student's $\bm{t}-$distribution with shape parameter $\nu$, scale parameter $b$, and mean equal to zero.}

For this $\bm{t}-$distribution the standard deviation equals $b \times \sqrt{(\nu/(\nu-2))}$, and thus the {standard deviation} is the product of a scale factor {(i.e., the volatility)} and a function that accounts for the fat tails of the distribution.  For large $\nu$ the Student's $\bm{t}-$distribution becomes a normal distribution, $\sqrt{(\nu/(\nu-2))}$ is approximately equal to unity, and the volatility is approximately equal to the standard deviation.  The dependence of the standard deviation on the volatility and a contribution from the shape of the underlying pdf might make estimation of the volatility from observations difficult.  It might be necessary to eliminate the maximum and minimum values from a small size sample in an estimation of the volatility \cite{Cas6}.

\begin{figure}[htbp] 
\centering
\includegraphics[scale = 0.7]{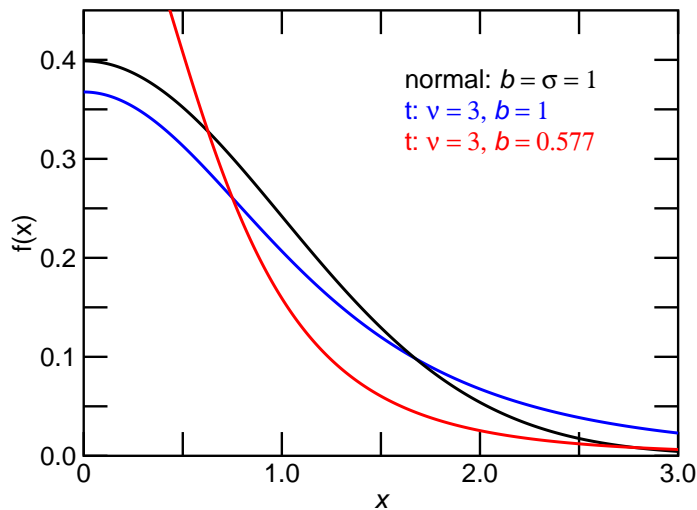} 
\caption{{\normalsize Comparison of a normal pdf, and $\nu=3$ Student's $\bm{t}-$distributions with $b=1$ and with $b=1/\sqrt{3}$.}}
\label{fig:fig4} 
\end{figure}

Figure~\ref{fig:fig4}  shows a comparison of a normal pdf with $\nu=3$ Student's $\bm{t}-$distributions with $b=1$ and $b=1/\sqrt{3}$.  For these three functions, the standard deviations equal 1, $\sqrt{3}$, and 1.  To compare prices based on Student's $\bm{t}-$distributions and normal distributions, the correct volatility must be used. 
\bigskip

\bigskip

\section{Data}
\begin{figure}[htbp] 
\centering
\includegraphics[scale = 0.7]{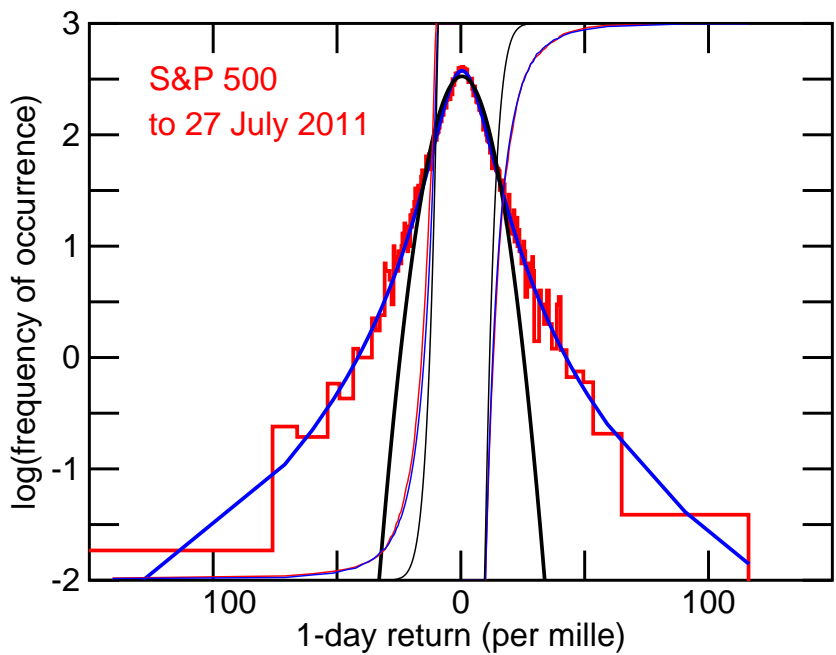} 
\caption{{\normalsize 1-day {linear} returns and fits of normal and Student's $\bm{t}-$distributions to the 1-day {linear} returns.}}
\label{fig:fig2a} 
\end{figure}

\begin{figure}[htbp] 
\centering
\includegraphics[scale = 0.7]{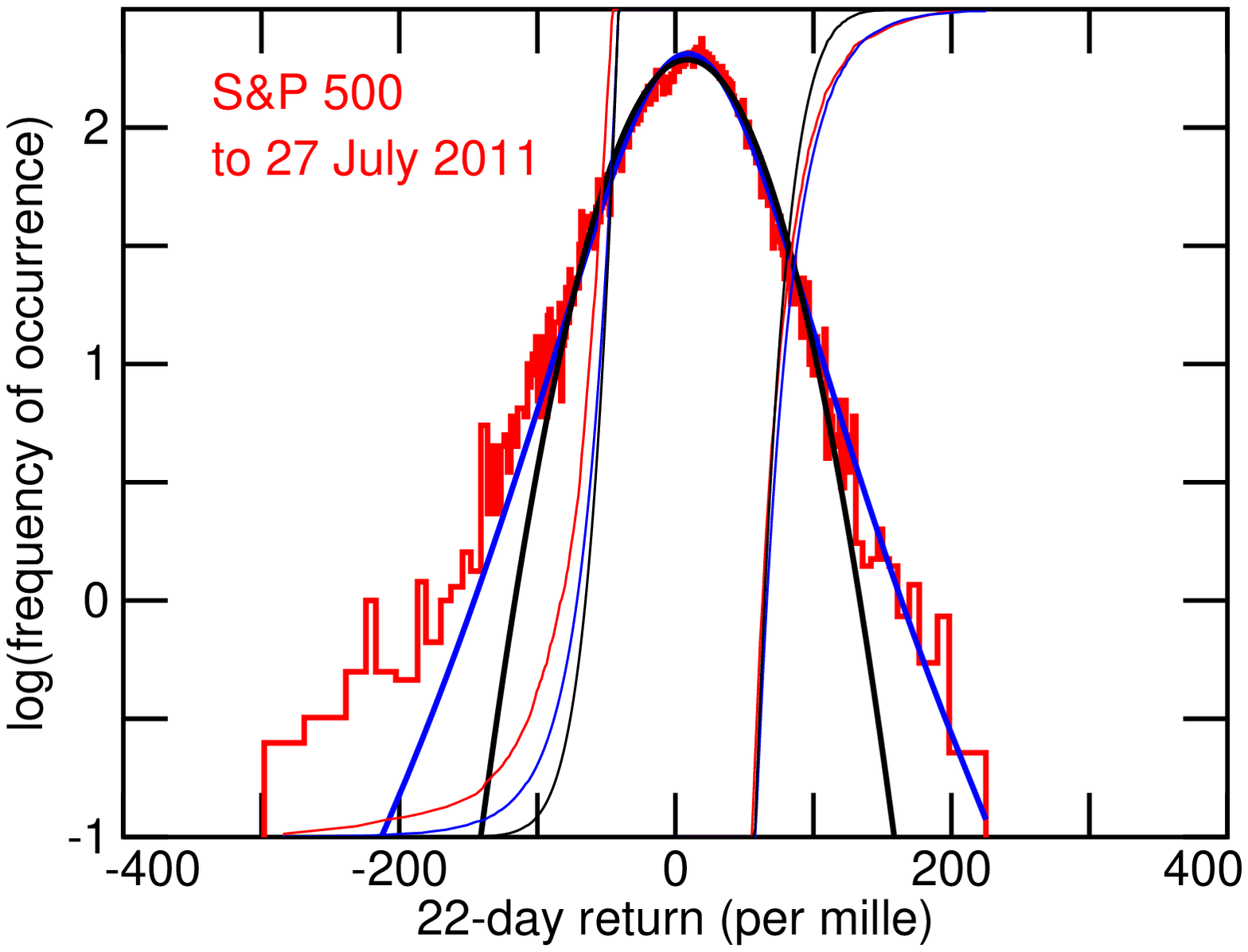} 
\caption{{\normalsize 22-day {linear} returns and fits of normal and Student's $\bm{t}-$distributions to the 22-day {linear} returns.}}
\label{fig:fig2b} 
\end{figure}

\begin{figure}[htbp] 
\centering
\includegraphics[scale = 0.7]{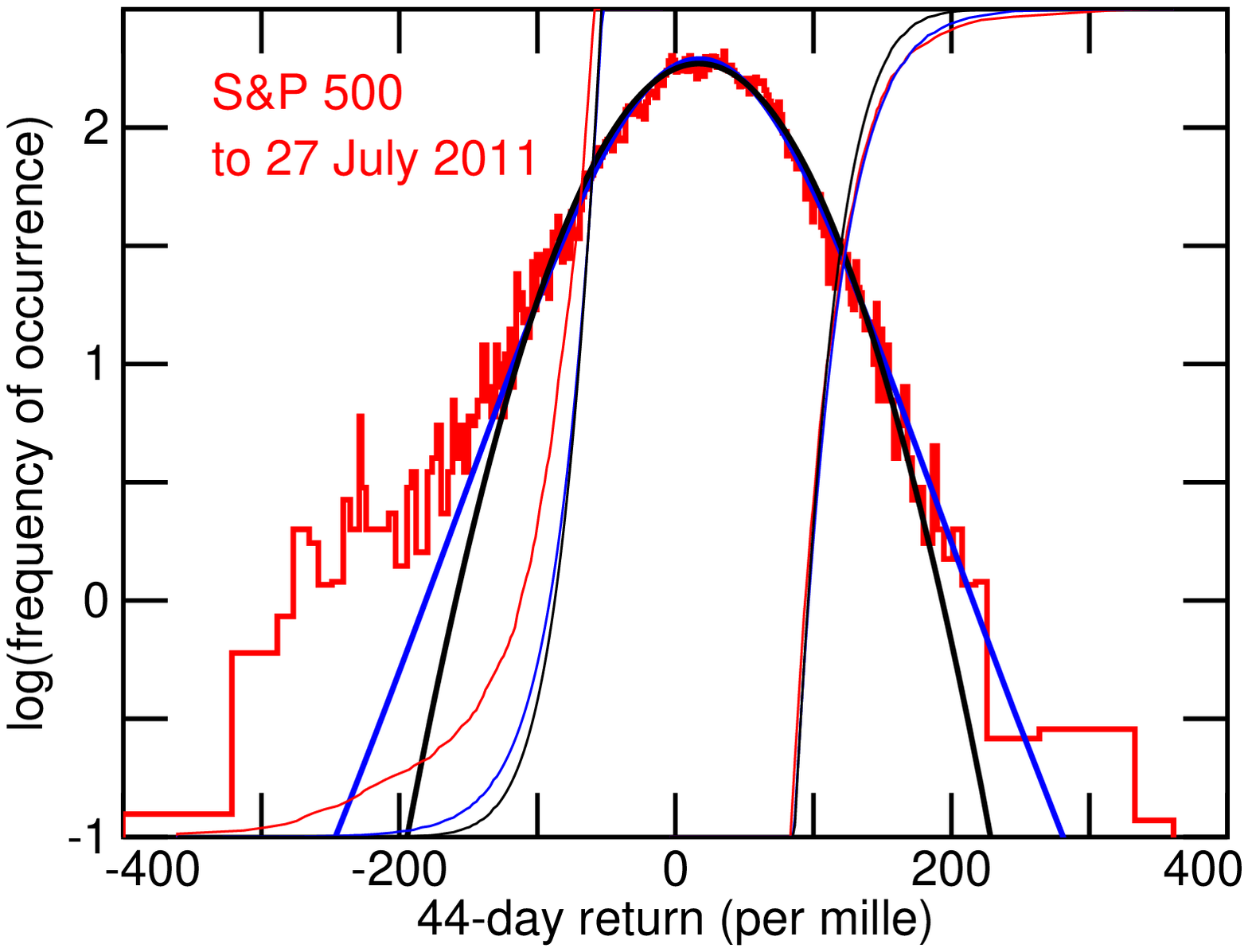} 
\caption{{\normalsize 44-day {linear} returns and fits of normal and Student's $\bm{t}-$distributions to the 44-day {linear} returns.}}
\label{fig:fig2c} 
\end{figure}

\begin{figure}[htbp] 
\centering
\includegraphics[scale = 0.7]{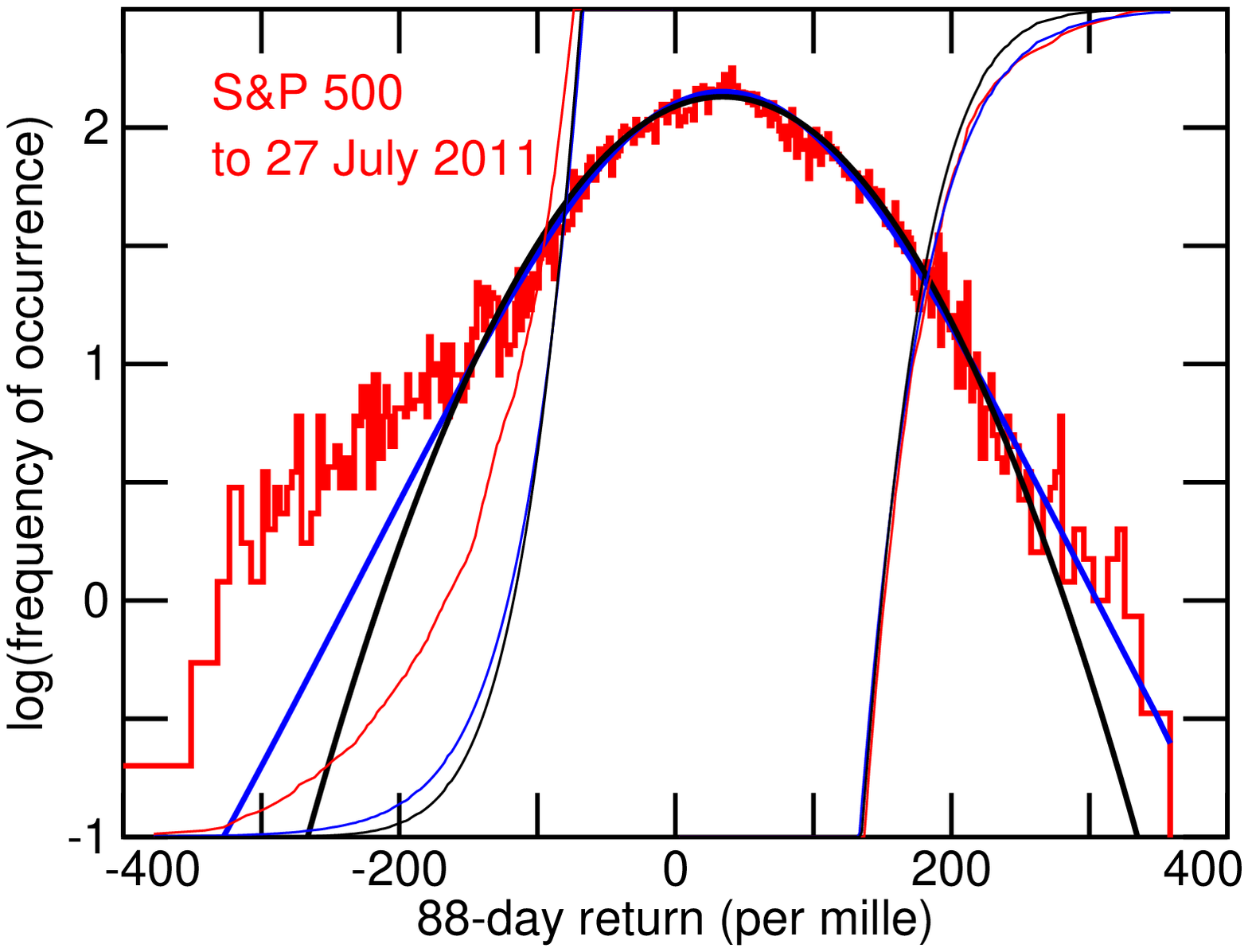} 
\caption{{\normalsize 88-day {linear} returns and fits of normal and Student's $\bm{t}-$distributions to the 88-day {linear} returns.}}
\label{fig:fig2d} 
\end{figure}

Figures~\ref{fig:fig2a}, ~\ref{fig:fig2b}, ~\ref{fig:fig2c} and~\ref{fig:fig2d} are semilog plots of the 1-day, 22-day, 44-day, and 88-day {linear }returns from the S\&P 500 index from 3 January 1950 to 27 July 2011 with best fit Student's $t$-distributions and best fit normal distributions.  The data are plotted as histograms without the vertical bars, hence the angular features in the plots of the data.  Adaptable bin widths were used, with the requirement that at least 5 counts occur in each bin.  The broad flat areas in the tails show the range of returns required to accumulate at least 5 counts.  The thin lines plot, on linear scales, the cumulative density functions (CDF) for the data (in red), the $\bm{t}-$distribution (in blue), and the normal distribution (in black).  Only the CDFs from 0 to 0.1 and from 0.9 to 1.0 are shown.  The CDF for the data and for a perfect fit should overlap.  The plots of the data, the fits, and the CDFs show that the data are fit well in the central region, with discrepancies in the tails.  The quality of the fit of the Student's $\bm{t}-$distribution to the 1-day returns is remarkable, as shown in Fig.~6.  

It is interesting to note that the prices of European call options require the values of the asset and the probability of the asset for $S(T) > K_T$ where $T$ is the expiration time of the asset and $K_T$ is the value of the strike at time $T$.  These values for the price of the asset and the probability are located on the right hand side of the plots, where the $\bm{t}-$distribution fits well the data.

\bigskip

Table~\ref{tbl2} gives descriptive statistics and best fit parameters for per mille {linear} returns for the S\&P 500 index from 3 January 1950 to 27 July 2011.
\begin{table}
\caption{Descriptive statistics and best fit parameters for per mille {linear} returns, S\&P 500 to 27 July 2011}
\begin{tabular}{c c c c c c} \hline
 & 1-day & 22-day & 44-day & 88-day & 128-day \\ \hline
count & 15491 & 15470 & 15448 & 15404 & 15364 \\ 
min & $-$205 & $-$298 & $-$400 & $-$403 & $-$469 \\ 
average & 0.329 & 7.22 & 14.4 & 29.1 & 42.9 \\ 
median & 0.463 & 10.4 & 17.8 & 32.6 & 46.3 \\ 
max & 116 & 224 & 360 & 358 & 527 \\ 
std dev & 9.67 & 44.2 & 62.4 & 89.5 & 113 \\ 
kurtosis & 25 & 6.1 & 5.8 & 4.3 & 4.0 \\ 
skewness & $-0.7$ & $-0.6$ & $-0.6$ & $-0.4$ & $-0.3$ \\ \hline
$\bm{t}$ &  &  &  &  &  \\ 
shape $(\nu)$ & 3.33 $\pm$ 0.2 & 6.3 $\pm$ 0.7 & 9.0 $\pm$ 1.3 & 12.6 $\pm$ 3 & 12.0 $\pm$ 3 \\ 
scale $(b)$ & 6.06 $\pm$ 0.3 & 34.7 $\pm$ 3 & 50.4 $\pm$ 4 & 73.9 $\pm$ 11 & 98 $\pm$ 11 \\ 
location & 0.46 $\pm$ 0.1 & 9.0 $\pm$ 0.6 & 16.9 $\pm$ 0.8 & 34.2 $\pm$ 1 & 45.8 $\pm$ 2 \\ \hline
normal &  &  &  &  &  \\ 
scale $(\sigma)$ & 7.3 $\pm$ 0.9 & 38.5 $\pm$ 0.4 & 54.5$\pm$ 0.9 & 79.2 $\pm$ 0.7 & 104.6 $\pm$ 1 \\ 
location & 0.42 $\pm$ 0.2 & 8.8 $\pm$ 0.7 & 17.1 $\pm$ 0.5 & 34.2 $\pm$ 1 & 46.5 $\pm$ 2 \\ \hline
\end{tabular}
\label{tbl2}
\end{table}

\bigskip
The uncertainties for the scale parameters are larger for fits to the Student's $\bm{t}-$distributions than for the fits to the normal distributions.  The fits to the $\bm{t}-$distributions are not inferior to fits to the normal distributions.  The shape and scale parameters are correlated in fits to the $\bm{t}-$distributions.  For a range of values it is possible to decrease one of the two correlated parameters{, decrease} the other, and still maintain a good fit. 
\bigskip

\begin{figure}[htbp] 
\centering
\includegraphics[scale = 0.7]{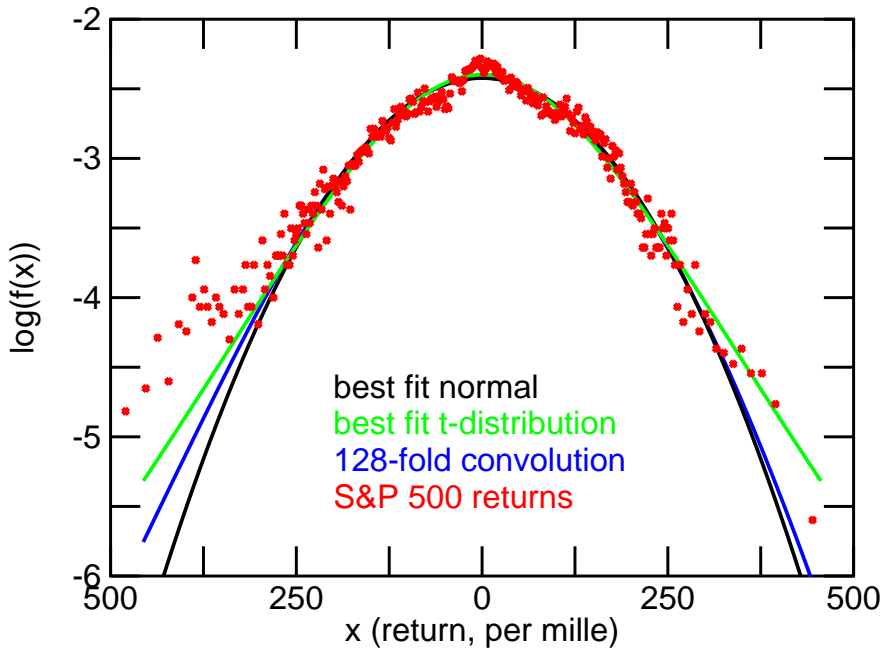} 
\caption{{\normalsize Best fits to the 128-day {linear} returns, data points, and 128-fold self-convolution.}}
\label{fig:fig7} 
\end{figure}

\bigskip

Figure \ref{fig:fig7} is comprised of plots of the best fit normal pdf, best fit Student's $\bm{t}-$distribution, the data points for the 128-day {linear} returns for the S\&P 500 data minus the average return, and the 128-fold convolution of the best fit Student's $\bm{t}-$distribution to the 1-day {linear} return for the S\&P 500 data.  The convolution was obtained by truncating the best fit $\bm{t}-$distribution to the 1-day {linear} returns for $\bm{t} < -305$ and $\bm{t} > 116$.  The minimum and maximum 1-day {linear} returns were $-205$ and $116$.  The asymmetric truncation of the $\bm{t}-$distribution leads to a slight asymmetry of the 128-fold convolution.   This asymmetry can be observed by comparing the difference between the left and right tails of the best fit normal and the 128-fold convolution.  The left truncation point of $-305$ was chosen to add some asymmetry, as the data show.  Figure \ref{fig:fig7} shows that the 128-day return is well approximated by the 128-fold convolution of the 1-day return.  No additional fitting parameters were used to obtain the 128-day return from the 128-fold convolution of the best fit 1-day Student's $\bm{t}-$distribution, Fig. \ref{fig:fig1}.  A better fit can be obtained by starting with a Student's $\bm{t}-$distribution with a smaller shape parameter $\nu$ than the shape parameter obtained from the best fit to the 1-day returns.  A smaller $\nu$ broadens the tails for the 128-fold convolution.  The best fit distribution parameters were used to demonstrate that the $n$-day return is to a good approximation the $n$-fold convolution of the 1-day return.  A smaller $\nu$ can be justified on the basis that the 1-day return is the convolution of the tic-by-tic returns.  The tic-by-tic returns will be described by distributions with small shape parameters, i.e., have fat tails, which when convolved to obtain the distribution for the daily adjusted closing value will be described by distributions with larger shape (or  {smaller }degrees of freedom) parameters.  The best fit Student's $\bm{t}-$distribution to the 1-day returns provides a convenient starting point.  The shape of a repeatedly self-convolved Student's $\bm{t}-$distribution and the rate that a repeatedly self-convolved Student's $\bm{t}-$distribution approaches a normal pdf are dependant on the truncation and the shape parameter~\cite{Cas3}.
\bigskip

\section{Prices of European Call Options}

Table~\ref{tbl3} gives $C_o$, the price of an $n$-day European call option, as determined from the solution $S(t)$ to the homogeneously saturated equation, Eq. (2), for noise driving terms $f(t)$ that follow a Student's $t$-distribution (top half of the table) or normal pdf (bottom half of the table).  For comparison $C_o$ as predicted by the Black-Scholes equation is given in the right hand column.  For all the pricing, the best fit parameters obtained for the fits to the S\&P 500 data were used.  These best fit parameters are given in the bottom portion of Table~\ref{tbl2} in Sec.~3. To calculate prices, the following values were assumed: the price of the asset at time zero $S_o=\$50.00$; the strike price at time $T = n$ days $K_T = \$49.00$; one year was taken as 252 days; and, the risk free rate $r=0.03$ per year. 

The expectation E$\{\max(S_T - K_T,0)\}$ was calculated by numerical integration to obtain $C_T = \exp(r\times T/252)\,C_o$ for a $T$-day option.  Monte Carlo simulation yielded similar results.  The pdf was divided into 1000 sections of equal area between $10^{-3}$ and 0.999, and was evaluated {at areas} of $1.0\times10^{-5}$, $2.5\times10^{-5}$, $5.0\times10^{-5}$, $1.0\times10^{-4}$, $2.5\times10^{-4}$, and $5.0\times10^{-4}$ and the symmetrical points near unity.  {Except for values beyond 0.99999, the required values of $S_T(x)$ were found from numerical solution to Eq. (2).}  For contributions from values beyond 0.99999, a linear approximation $S_T(x) = S_T(0) + q\,x/\beta$ was multiplied by the analytic expression for the pdf  {for $S_T$} and integrated to infinity.  $q$ was the input scale parameter reported in Table~\ref{tbl1}.  The pdf for $S_T$ was obtained by {using} the  {defining} equation for the $T$-day {linear} return $R_T$, $R_T= 1000 \times (S_T/S_o - 1)$, and transforming the {best-fit} pdf  {for $R_T$}.  Trapezoidal integration was employed {to determine the expectation.  Here the notation $S_T(x)$ indicates that the price of the asset $S(t)$ is to be evaluated for $\alpha t + W(t) = x$ and at an expiry time $T$.  The value of the subscript $T$ has no bearing on the solution to the price.  The subscript $T$ indicates the shape of the pdf to associate with the price.}

The values for $C_o$ in Table~\ref{tbl3} show a dependence on $\beta$, with $C_o$ decreasing with an increase of $\beta$ and the effect becoming more pronounced as $n$ of $n$-day increases.  These dependencies on $\beta$ and $n$ owe to noise rectification.  The values in the tail of the distribution push the solution into the non-linear region.  This gives a non-zero value to the output noise (the input noise has a mean value of zero) and this increases the mean value of the solution.  The noise rectification was not controlled in the calculations reported in Table~\ref{tbl3}.  The amount of rectification increases as the width of the input distribution increases.  This explains the increase with $n$.  {For $T=128$ days, the mean value for $S(T)$} was calculated as \$51.374 for $\beta=0.001$ and \$51.098 for $\beta=3.0$ for a noise driving term that follows a Student's $\bm{t}-$distribution.  For normally distributed noise, the mean values {for $S(T)$} were found to be \$51.031 and \$50.770.  The true value was $S_o\exp(0.03\, T/252) = \$50.7677$.

Values for $C_o$ calculated using the best-fit $\bm{t}-$distributions are greater than the prices found from the normal pdf and from the Black-Scholes equation.  The values of $C_o$ found for the normal distribution using Eq. (2) are similar to the values found from the Black-Scholes equation.  In the Black-Scholes equation, the stock price is assumed to follow a log-normal distribution.  For the normal distribution and the homogeneously saturated equation, the stock price follows a normal distribution.  Pinn \cite{Pinn} considered pricing of options when the distribution of the prices of the underlying stock followed a Student's $\bm{t}-$distribution. 

\begin{table}
\caption{$C_o$ from the solution $S(t)$ to Eq. (2) for noise drawn from a Student's $\bm{t}-$distribution and from a normal distribution.  {For comparison, $C_o$ as determined from the Black-Scholes equation is displayed.  $S_o=\$50.00$ and $K_T=\$49.00$. }}
\begin{tabular}{c c c c c c c c } \hline
 & $\beta = 0.001$ & $\beta = 0.01$ & $\beta = 0.1$ & $\beta = 0.3$ & $\beta = 1.0$&$\beta = 3.0$& Black Scholes \\ \hline
$\bm{t}$ &  &  &  &  &  & &\\ 
1-day & $1.025$ & $1.020$ & $1.018$ & $1.018$ & $1.018$ & $1.018$& \\ 
22-day & 1.527 & 1.516 & 1.508 & 1.505 & 1.504 & 1.504 &  \\ 
44-day & 1.910 & 1.895 & 1.869 & 1.865 & 1.863 & 1.862 &   \\ 
88-day & 2.543 & 2.520 & 2.476 & 2.467 & 2.466 & 2.463 & \\ 
128 day & 3.242 & 3.196 & 3.125 & 3.109 & 3.102 & 3.100 & \\ \hline
normal &  &  &  &  &  & &\\ 
1-day & $1.008$ & $1.007$ & $1.007$ & $1.006$ & $1.006$ & $1.006$&1.0061\\ 
22-day &1.488 & 1.481 & 1.469 & 1.467 & 1.466 & 1.466 & 1.4529 \\ 
44-day & 1.874 & 1.861 & 1.840 & 1.836 & 1.835 & 1.834 & 1.8162\\ 
88-day & 2.531 & 2.508 & 2.472 & 2.459 & 2.455& 2.455 & 2.4261\\ 
128-day & 3.207 & 3.167 & 3.104 & 3.090 & 3.083 & 3.082 & 3.0361 \\ \hline
\end{tabular}
\label{tbl3}
\end{table}

Values for $C_o$ were also found for numerical integration of the S\&P 500 returns, in contrast to numerical integration of the best-fit distribution.   These values are listed in Table~\ref{tbl4} along with the Black-Scholes prices.  The numerical integration of the S\&P 500 data leads to larger values for the price of the option {($S_o=\$50.00$, $K_T = \$49.00$; $r=0.03$ per year, best-fit volatility)}  than is obtained from integration of the best fit {distributions}.

\begin{table}
\caption{$C_o$ from numerical integration of S\&P 500 data and from the Black-Scholes equation.  $S_o=\$50.00$ and $K_T=\$49.00$.}
\begin{tabular}{c c c c c c} \hline
 & 1-day & 22-day & 44-day & 88-day & 128-day \\ \hline
S\&P data &  &  &  &  &  \\ 
$C_o$ & 1.163  & 1.647 & 2.243 & 2.642 & 3.509 \\ \hline
Black Scholes &  &  &  &  &  \\ 
$C_o$ & 1.006  & 1.453 & 1.816 & 2.426 & 3.036 \\ \hline
\end{tabular}
\label{tbl4}
\end{table}

Table~\ref{tbl5} gives values of $C_o$ for values of the strike $K_T$ for the homogeneously broadened equation with $\bm{t}$ and normal statistics, for the Black-Scholes equation, for direct numerical integration of the S\&P 500 data, and for a Gosset~\cite{Cas1} formula.  For these calculations, $\beta=0.3$, $S_o=50.0$, $r =0.03$, and $n=22$.  The Gosset formulae price European call options when the underlying stock price is distributed as a log-Student's $\bm{t}-$distribution.  The Gosset formulae truncate the underlying pdf or cap the value of the stock to keep the integrals involved in the pricing finite.  For the calculations presented in Table~\ref{tbl5} the $\bm{t}-$distribution was truncated at $p=0.9999$.   For the 22-day best fit parameters of $\nu=6.3$ and $b = 34.7$, this yields truncation of per mille returns of $> 266$.  From Table~\ref{tbl2} the maximum 22-day per mille return was 224, which includes an average return of 7.  Thus the truncation for the Gosset formula is reasonable.
 
\begin{table}
\caption{$C_o$ for different strike prices and methods of calculation with $T = 22$ days, $\beta=0.3$, and $S_o=\$50.00$.}
\begin{tabular}{c c c c c c} \hline
  $K_T$& $\bm{t}$ & normal & Black-Scholes & S\&P data & Gosset\\ \hline
40.0 & 10.120 & 10.107  & 10.105  & 9.830 & 10.11  \\ 
42.5 & 7.626  & 7.614 & 7.611 & 7.427 & 7.613 \\ 
45.0 &  5.144 & 5.122 & 5.119 & 5.075 & 5.131   \\ 
47.5 & 2.743  & 2.702 & 2.693 & 2.822 & 2.726  \\ 
50.0 & 0.861  & 0.836 & 0.834 & 1.031 & 0.856  \\ 
52.5 & 0.141  & 0.103 & 0.111 & 0.276 & 0.150  \\ 
55.0 &  0.022 & 0.004 & 0.005 & 0.108 & 0.026  \\ 
57.5 &  0.004 & 0.000 & 0.000 & 0.035 & 0.005  \\ 
60.0 &  0.001 & 0.000 & 0.000 & 0.002 & 0.001  \\ \hline
\end{tabular}
\label{tbl5}
\end{table}

The five methods predict similar trends and results, with the exception of the direct numerical integration of the S\&P 500 data.  The numerical integration of the S\&P data gives a decreasing price for $K_T \le \$45.0$.  $K_T = \$45.0$ maps to a per mille return of $-100$.  From Fig.~{7}, it can be observed that the calculation of the expectation required for the price starts to access the left hand shoulder on the 22-day return.  This will bias the calculation towards smaller values of $S_T-K_T$ and reduce $C_o$.

\bigskip

\section{Conclusion} 
 A homogeneously saturated equation that gives the time development of the price of an asset was presented and used to price European call options.  The options were priced using best-fit Student's $\bm{t}-$distributions and best-fit normal distributions to {linear returns calculated from the} daily adjusted closing values for the S\&P 500 Index from 3 January 1950 to 27 July 2011.  The prices as predicted by the homogeneously saturated equation were compared to prices obtained from direct integration of the S\&P 500 data, from the Black-Scholes equation, and from a Gosset formula, and were found to be in agreement.
 
 The homogeneously saturated equation borrows from laser physics.  The homogeneously saturated equation is similar to the equation for the output of a simple, homogeneously broadened laser.  The homogeneously saturated equation for the time development of the price of an asset was justified by solution in steady state of coupled rate equations that describe a reservoir of money that can be used to purchase an asset and the stock price.  A noise driving term was added to the rate equation for the reservoir of money.  The equations thus describe the effect of a fluctuating money supply on the price of the asset.

The homogeneously saturated equation for the time development of the price of an asset has some interesting and realistic features.  In the limit that a saturation parameter $\beta$, which is a measure of the strength of the coupling of the reservoir and price of the asset, equals zero, the standard equation for the time development of the price of an asset is obtained.  The solution to the standard equation is geometric motion, or geometric Brownian motion, if the noise forcing term $\sigma f(t)$ follows normal statistics.  

The solution to the homogeneously saturated equation tends, through the saturation inherent to the form of the coupled equations, to depend linearly on the input to the model for input $x$ such that $S(x) \gg 0$ and to compress the output.  Fluctuations in the input to the homogeneously saturated equation create smaller fluctuations in the solution for non-zero $\beta$.  The transfer function of the homogeneously saturated equation goes as $x/\beta$ where $x$ is the input to the equation for $x$ such that $S(x) \gg 0$.  This is in contrast to the standard model, where the output $S(t) = \exp( \int \sigma f(t)\, \rm{d}t)$ depends exponentially on the input.  With the standard model and with a noise forcing term (i.e., the input) that follows fat tailed statistics (such as the Student's $\bm{t}-$distribution, which is known to fit well the returns), integrals that are required to price European call options are infinite .  The homogeneously saturated equation, through linearization and compression by saturation of the reservoir, does not suffer the same shortcoming as the standard model.  In this respect the homogeneously saturated equation for the development in time of the price of an asset corresponds to reality better than the standard model, and thus should be preferred over the standard model for the development in time of the price of an asset.

\section{Appendix} 
In this Appendix, rate equations for a reservoir of money $M(t)$ that is available to invest in an asset with price $S(t)$ and for the time development of the price of the asset are constructed.  The rate equations are written as Langevin equations, which are first order differential equations with noise driving terms.  The Langevin equations should be interpreted as integral equations  \cite[pg 172]{Cof} \cite[Ch 10.2]{Lax}.  Average values found by the Langevin approach are identical to solutions found by Ito's calculus \cite{Lax}.

Let {\it M\/}({\it t\/}) be the amount of money that is available to invest in an asset.  Let {\it N\/} be
the rate at which money is pumped into the reservoir of money {\it M\/}({\it t\/})  that can be
used to purchase the asset and let $\beta/\tau\times${\it M\/}({\it t\/})$\times${\it S\/}({\it t\/}) be the rate that money is removed
from the reservoir owing to purchases of the asset.  Let $\tau$ be a characteristic time
constant that allows for money to be removed or added to the reservoir, depending
on whether {\it M\/}({\it t\/}) is greater than or less than some value {\it M$_{{\rm o}}$\/}.  Allow for a noise driving term $\sigma/\tau f(t)$. 

A rate equation for {\it M\/}({\it t\/}) is then


\begin{equation}
{\frac{\rm d}{{\rm d} \, t}} M(t)~=~N~-~\frac{\beta}{\tau} \times  S(t) \times  M(t)\, ~-~{\frac{M(t)~-~M_{o} }{\tau }} + \frac{\sigma}{\tau} f(t)\,\,.
\end{equation}

All parameters in the rate equation have a time dependence, but it is assumed that  these parameters change slowly in time.  Thus each point in time is assumed to evolve about a steady state .  In steady state, the time derivative equals zero and 
\begin{equation}
M(t) ~=~{\frac{N ~+~{\frac{~M_{o} }{\tau }} ~+~\frac{\sigma}{\tau} \, f\, (t)}{{\frac{1}{\tau }} ~+~\frac{\beta}{\tau} \, S(t)}} ~=~{\frac{(\alpha ~+~\sigma \, f\, (t))}{1~+~\beta \, S(t)}} {\rm ~.}
\end{equation}

In the last form for {\it M\/}({\it t\/}) the symbol $\alpha$ has been defined.  The equation for the time development of the value of an asset becomes 

\begin{equation}
{\frac{\rm d}{{\rm d} \, t}} S(t)~=~M(t)\, S(t)~~=~{\frac{\alpha \, S(t) ~+~\sigma \, S(t)\, f\, (t)}{1~+~\beta \, S(t)}} \,\,.
\end{equation}

In this approach the noise is ascribed to fluctuations in the amount of money available to invest in the asset.  The value at time $t$ of an asset under this homogeneously saturated approach is $S(t)$, the solution to Eq. (2).  

Note that the interaction between the reservoir and the rate that money left the reservoir was defined as $\beta/\tau$ where $\tau$ is a characteristic time constant of the reservoir.  This definition defines $\beta$ in terms of a characteristic time $\tau$ of the system and suggests that a reasonable value for $\beta$ might be of order unity.

\bigskip
\section{References}


\end{document}